\documentclass[aps,prl,twocolumn]{revtex4}
\usepackage{amsfonts}
\usepackage{amssymb} 
\begin{document}
\title{$J^{PC}$-Exotic Mesons from the Bethe-Salpeter Equation} 
\author{C.~J.~Burden}
\affiliation{Department of Theoretical Physics,
Research School of Physical Sciences and Engineering,
Australian National University, Canberra, ACT 0200, Australia}
\author{M.~A.~Pichowsky}
\affiliation{Center for Nuclear Research, 
Department of Physics, Kent State University, Kent, OH 44242, U.S.A.}
\begin{abstract}
Masses for vector-$1^{--}$, axialvector-$1^{+-}$, $1^{++}$,
and exotic-vector $1^{-+}$ mesons are calculated within a quantum field
theoretic framework based on the Dyson-Schwinger equations using a
model form of the quark-antiquark interaction that is separable.
The model provides an excellent description of $\pi$ and
$K$-meson observables and the flavor-octet ground-state meson spectrum.
With no adjustments to model parameters, a numerical solution of the
Bethe-Salpeter equation yields two exotic-$1^{-+}$ mesons with masses of
$1439$ and $1498$~MeV.   
\end{abstract}
\maketitle
\section{Introduction}
The $\pi_1(1400)$ meson, which was until recently referred to as
$\hat{\rho}(1405)$, may be the first direct evidence of a non-$q\bar q$
meson.  With an estimated mass of $1376\pm 17$~MeV and width of $300\pm
40$~MeV \cite{PDG}, this resonance is distinguished by the {\em exotic}
quantum numbers $J^{PC} = 1^{-+}$.  Evidence of this resonance was observed
in $\pi p$ scattering by the E852 Collaboration~\cite{T97} at Brookhaven and
in $\bar{p} d$ annihilation by the Crystal Barrel Collaboration~\cite{A98} at
CERN.  Difficulties encountered in extracting information about this
resonance have raised some doubts as to its existence.  However, more
recently, a second observed exotic-$1^{-+}$ resonance has gained wider
acceptance.  This second resonance was observed by the E852
Collaboration~\cite{Adams98} at the slightly higher mass of $1593\pm 8$ MeV
and with a width of $168\pm 20$ MeV.  With the construction of a new
experimental Hall D and an energy upgrade proposed for the Thomas Jefferson
National Accelerator Facility, the exotic meson spectrum will be probed with
an accuracy hitherto unknown; the data gathered should provide compelling
evidence, either for or against, the existence and masses of these two
$J^{PC}$-exotic states.

Mesons with so-called {\em exotic} quantum numbers $J^{PC} = 0^{--}$,
$0^{+-}$, $1^{-+}$, $2^{+-}$, etc., are traditionally interpreted within a
quantum mechanical framework as either {\em hybrids} or {\em molecules}; that
is, $q \bar{q} g$ or $q\bar{q}q\bar{q}$ states, respectively \cite{P97}.  The
conventional quantum mechanical picture of a meson is that it is comprised of
a quark-antiquark {\em state} $|q\bar{q}\rangle$ and so must have a
space-inversion parity of $P = (-1)^{L + 1}$, and a charge-conjugation parity
of $C = (-1)^{L + S}$, where $L$ is the relative orbital angular momentum and
$S$ is the total spin of the system.  Therefore, it is impossible to
construct a quark-antiquark state that has $J^{PC}=1^{-+}$ quantum numbers.
Analogously, lattice-QCD simulations of the exotic meson spectrum employ
operators that contain explicit excitations of the gluon degrees of freedom.
This is done either by joining quark sources by link operators corresponding
to transverse excitations of the color flux tube~\cite{LMBR96} or by
combining plaquette excitations with quark bilinears~\cite{Be97}.

In what may appear to be in contradiction to the aforementioned approaches,
one finds that in a fully-covariant quantum field theoretic treatment of
meson bound  states, one can in fact construct quark bilinear operators
which have $J^{PC}$-exotic quantum numbers without the introduction of
{\em explicit} gluon degrees of freedom. 
This is a result of having an additional degree of freedom available in
quantum field theory arising from the relative-time between the quark and
antiquark, leading to an additional factor of $(-1)^{\kappa}$ in the
charge-conjugation parity relation $C = (-1)^{L+S+\kappa}$, 
where $\kappa$ is the relative-time quantum number analogous to $L$.
The usual even-time parity operators employed within studies of non-exotic
mesons have  $\kappa = 0, 2, 4,\ldots$.  
Operators for which $\kappa$ is an odd integer are referred to as
``odd-time parity'' operators. 
Although the existence of two-body bound-state solutions with
$J^{PC}$-exotic quantum numbers has been known since the earliest studies
of the Bethe-Salpeter equation \cite{N65}, the details of their role and
interpretation within quantum field theory remain unresolved.

It should be emphasized that the use of such odd-time-parity operators,
which are devoid of explicit gluon degrees of freedom, to generate exotic
meson solutions from the BS  equation does  {\em not} imply that exotic
mesons have a non-zero overlap with the state $|q\bar{q}\rangle$.
Rather, since such operators have $\kappa \not = 0$, they vanish when
the relative time between the quark and antiquark is zero, and hence
have {\em no} overlap with a $|q\bar{q}\rangle$ state.  
That is, the states with the fewest number of particles that could appear
in a Fock space expansion of an exotic meson would be states such as 
$|\bar{q}q\bar{q}q\rangle$ and $|\bar{q}qg\rangle$.
This is, of course, consistent with the convention picture described
above. 

In principle, one could also employ similar quark bilinear operators 
in lattice-QCD simulations.
Some examples of suitable operators for the $1^{-+}$-exotic meson would be
the lattice versions of $\bar{\psi}\gamma_i D_4\psi$ and
$\bar{\psi}\gamma_4 D_i\psi$ which  are constructed using the
gauge-covariant derivative  $D_{\mu} = \partial_{\mu} + ieA_{\mu}$.   
In a framework, invariant under local SU(3)-color transformations,
the use of such covariant derivatives inextricably mixes 
odd-time parity quark bilinear operators with the usual 
quark-antiquark-gluon hybrid states considered in other models.
However, in practice lattice-QCD simulations are typically carried out at
quark masses greater than the strange quark mass and hence are
uncomfortably close to the non-relativistic limit.  
Because odd-time-parity operators go to zero in the limit of infinitely
heavy quarks, their use in lattice-QCD calculations is not promising as
any signal is likely to be swamped by noise~\cite{LMBR96}. 
In contrast to lattice gauge theory, this poses no difficulty in the
present study, as the Dyson-Schwinger framework has proved to be directly
applicable to 
the limit of small current quark masses.  Hence, odd-time-parity
operators may provide a useful tool for the study of $J^{PC}$-exotic mesons
within this framework.

In the present study of the $1^{-+}$-exotic spectrum, 
we employ a quantum field theoretic framework 
based on the Dyson-Schwinger and Bethe-Salpeter (BS)
equations~\cite{RobertsWilliams}.   
The use of quark bilinear, odd-time parity operators with $J^{PC}$-exotic
quantum numbers allows one to obtain solutions of the homogeneous
quark-antiquark BS equation that have non-zero overlap with
$J^{PC}$-exotic mesons.   
In this way, we obtain mass predictions for exotic mesons without the
{\em explicit} introduction of gluon degrees of freedom.

\section{Bethe-Salpeter Equation for Vector and Axialvector Mesons}
Our starting point is a calculation of the ground state spectrum of
light-quark mesons in which a confining, separable Ansatz for the BS
kernel is constructed from phenomenologically efficacious, confined quark  
propagators~\cite{BQRTT97}.  
The model was used to calculate the masses for the ground state
flavor-octet scalar, psuedoscalar, vector, and $J^{PC}=1^{-+}$ axialvector
mesons with remarkable success.
The authors also explored the possibility of exotic-scalar and
exotic-pseudoscalar states.  For the lightest of these mesons with
$J^{PC}=0^{+-}$ and $0^{--}$ quantum numbers, they obtained masses of 1082
and 1319~MeV, respectively.  The value for exotic-pseudoscalar meson seems
plausible, although there is no experimental consensus regarding the mass of
such a state at present.  However, the mass of the exotic-scalar meson
appears too light.
This may be the result of a known weakness of ladder-like Bethe-Salpeter
kernels in the study of scalar mesons~\cite{truncscheme}.  That weakness is
evident in Ref.~\cite{BQRTT97} through the predicted mass of the $0^{++}$
mesons: $749$~MeV, which is too light to be satisfactorily identified with
the $f_0$ and $a_0$, but too heavy for the putative $\sigma$-meson.
On the other hand, the results in Ref.~\cite{BQRTT97} are robust when it
comes to predicting masses of pseudoscalar, vector and axialvector mesons,
typically giving results within a few percent of the experimental values.
This outcome can also be anticipated~\cite{will}.
If this pattern is maintained in exotic channels, one expects that the mass
obtained in Ref.~\cite{BQRTT97} for the exotic-scalar meson is unreliable,
while masses obtained for the exotic-pseudoscalar channel, as well as the
exotic-vector meson channel considered in this study, would be dependable.

Before proceeding, one should note that exotic-meson solutions of the BSE
do not arise as a result of the use of a separable-kernel Ansatz, like the
one employed in Ref.~\cite{BQRTT97}.
Such solutions have appeared in many different Bethe-Salpeter
calculations and are also expected to arise in more sophisticated models
that use local, dressed-gluon exchange kernels, such as the very successful
model considered in Refs.~\cite{MarisTandy}.

The quark propagators used herein are based on numerical solutions of model
quark Dyson-Schwinger equations, and are algebraic functions fit to a broad
range of pion and kaon observables, including the decay constants $f_\pi$ and
$f_K$, charge radii $r_\pi$ and $r_{K}$, $\pi$-$\pi$ scattering lengths, the
pion and kaon electrodynamic form factor $F_{\pi}(Q^2)$ and $F_{K}(Q^2)$, and
the anomalous $\gamma\pi\rightarrow\gamma$ and $\gamma\pi\rightarrow\pi\pi$
transition form factors~\cite{R96etc}.  Quark confinement is explicitly
included in this model by the fact that the quark propagator $S(p)$ is
analytic for all finite values of the quark momentum $p$ on the complex-$p$
plane.  {}From this, it follows that this quark propagator has {\em no}
Lehmann representation and describes the propagation of a confined quark
\cite{RobertsWilliams}.

The calculation of meson bound states consists of solving the
quark-antiquark BS equation in the dressed-ladder approximation
\begin{eqnarray}
\Gamma_{\alpha}(p,P) &\!=\!& - \frac{4}{3} \!\int\!\! \frac{d^4q}{(2\pi)^4} 
     \,g^2 D_{\mu\nu}(p - q) 
\gamma_\mu S(q + \mbox{$\frac{1}{2}$} P)  
 \nonumber \\ && \times
\Gamma_{\alpha}(q,P) S(q - \mbox{$\frac{1}{2}$} P)
   \gamma_\nu~, 
\label{bsv} 
\end{eqnarray}
where $\Gamma_{\alpha}(q,P)$ is the BS amplitude, 
$S(p) = -i\gamma\cdot p \, \sigma_V(p^2) + \sigma_S(p^2)$ 
is the Euclidean-space confined-quark propagator for $u$ and $d$ quarks, 
\begin{equation}
K(p,q) = -\frac{4}{3}g^2\gamma_\mu D_{\mu\nu}(p - q)\gamma_\nu,
\label{kernel}
\end{equation}
is the dressed-ladder approximation to the quark-antiquark scattering kernel, 
and the hermitian Euclidean Dirac matrices $\gamma_{\mu}$ satisfy
$\{\gamma_\mu,\gamma_\nu\}=2\delta_{\mu\nu}$.  
By choosing for the kernel $K(p,q)$, the separable-gluon Ansatz  
\begin{equation}
\label{sepeqm}
g^2 D_{\mu\nu}(p-q) = \delta_{\mu\nu}\left[
G(p^2)\,G(q^2)\,+\,p\cdot q\,F(p^2)\,F(q^2) \right]~, 
\end{equation}
the Dyson-Schwinger equation for the quark propagator can be inverted to
give an 
algebraic relation between the kernel functions
$F(p^2)$ and $G(p^2)$ and the Lorentz-invariant functions
$\sigma_V(p^2)$ and $\sigma_S(p^2)$ in the quark propagator $S(p)$.    
The model kernel is completely determined once the two functions
$\sigma_V(p^2)$ and $\sigma_S(p^2)$ are specified.
These functions are fit to a range of pion and kaon observables in 
Ref.~\cite{BQRTT97} and have been successfully applied to study a plethora
of other meson reactions \cite{R96etc}.
The resulting solutions of the BS equation using the separable kernel of
Eq.~(\ref{sepeqm}) provide an excellent description of the ground state
flavor octet pseudoscalar, vector and axialvector meson
spectrum~\cite{BQRTT97}.  

Restrictions placed on BS amplitudes by Lorentz covariance, space
inversion, charge conjugation and the separable Ansatz of Eq.~(\ref{sepeqm}), 
imply that the most general forms of the transverse components
$\Gamma_i$, $i = 1$, 2, 3 of the $J^{PC} = 1^{--}$, $1^{+-}$, $1^{++}$ and
$1^{-+}$ amplitudes can be obtained from~\cite{BQRTT97}  
\begin{eqnarray}
\Gamma^{--}_i(q,P) & = &  q_i  F(q^2)\lambda_1 + 
      	i \gamma_i G(q^2) \lambda_2   
 \nonumber \\  & & \mbox{}
	+\gamma_5\epsilon_{ijk}\gamma_j q_k  F(q^2)\lambda_3,
                                  \label{Gi--} \\
\Gamma^{+-}_i(q,P) & = & 
    	i\gamma_5\gamma_i q_4 F(q^2) \lambda_1 + 
  	i\gamma_5 q_i F(q^2) \lambda_2 
\nonumber \\ & & \mbox{}
	+ i \gamma_5 q_i \gamma_4 F(q^2) \lambda_3,  
                            \label{Gi+-} \\
\Gamma^{++}_i(q,P) &=& i\gamma_5\gamma_i G(q^2) \lambda_1 + 
   	\epsilon_{ijk}\gamma_j q_k F(q^2)\lambda_2 ,      \label{Gi++} \\
\Gamma^{-+}_i(q,P) &=& \gamma_i q_4 F(q^2) \lambda_1 +
     	q_i  \gamma_4 F(q^2)\lambda_2, \label{Gi-+}  
\end{eqnarray}
where the coefficients $\lambda_A$ are obtained by solving the BS
equation for $P^2 = -M^2$, with $M$ being the mass of the meson bound
state under consideration.  
The resulting coefficients $\lambda_A$ for each of the channels considered
in Eqs.~(\ref{Gi--})--(\ref{Gi-+}) are given, along with the corresponding
value of $M$, in Table~\ref{Tab}.  
The coefficients have been normalized using the formula Eq.~(\ref{norm})
discussed below.  

\begin{widetext}
\begin{center}
\begin{table}  
\caption{
Normalized amplitudes and bound-state masses obtained from
Bethe-Salpeter equation for axial, vector, and exotic-vector mesons are
compared with the observed masses from \cite{PDG}.
All masses are given in units of MeV.
}
\begin{tabular}{c|c|rrr|c|c}
\rule{0mm}{5mm}
$J^{PC}$& $M$ &$\lambda_1$&$\lambda_2$&$\lambda_3$ 
	& Identification & Observed Masses \\ \hline
 $1^{--}$ & 736 & $-$0.075 & 0.331 & $-$0.049 
	& $\rho(770)$, $\omega(782)$ & 770$\pm$1, 782$\pm$0 \\
 $1^{+-}$ & 1244 & 0.017 &    0.587 &  0.145 
	& $h_1(1170)$, $b_1(1235)$ & 1170$\pm$20, 1230$\pm$3 \\
 $1^{++}$ & 1337 & 0.056 & $-$0.276 &         
	& $a_1(1260)$, $f_1(1285)$ & 1230$\pm$40, 1282$\pm$1 \\
 $1^{-+}$ & 1439 & 0.299 & $-$0.053 &         
	& $\pi_1(1400)$ & 1376$\pm$17  \\
 $1^{-+}$ & 1498 & 0.040 & $ $0.251 &         
	& $\pi_1(1600)$ & 1593$\pm$8
\end{tabular}
\label{Tab}  
\end{table} 
\end{center}
\end{widetext}

The masses that result from our calculations for the lightest
vector, axialvector and two $1^{-+}$-exotic mesons are 
compared with the experimental values in Table~\ref{Tab}. 
The masses obtained for the non-exotic vector and axialvector mesons are
each within 10\% of their accepted values.
The results for the $J^{PC} = 1^{--}$ ($\rho$/$\omega$) and $1^{++}$
($a_1$/$f_1$) have been published previously~\cite{BQRTT97}, 
and recently, calculations of the masses and decays of the $1^{+-}$
$h_1(1170)$ and $b_1(1235)$ mesons have also appeared~\cite{Roberts}.  
The new results of the present study are the two solutions of the BS
equation with $1^{-+}$-exotic quantum numbers.  The first has a mass of
1439~MeV and the second has a mass of 1498~MeV. 
These results are truly predictions of the model in that none of the
parameters determined in Ref.~\cite{BQRTT97} have been adjusted for the
calculations of any of the mesons obtained here.

It is interesting that the mass predictions obtained herein for the 
exotic-vector meson are somewhat lower than the range 1700--2000~MeV
obtained within quantum mechanical models \cite{P97} and lattice-QCD
simulations \cite{McN99}.   
Our results are in the range of mass values of 1300--1700~MeV obtained
from bag model calculations \cite{BC82} (although these values are
believed to suffer from parameter uncertainties \cite{DK99}), and in
the large range of values obtained using QCD sum rules \cite{BDY86} for
which the lightest mass for exotic-$1^{-+}$ mesons varies between
1300--2100~MeV.    

Our calculations suggest that there should be two $1^{-+}$-exotic mesons
in close proximity to each other.  If this is true, then one expects the
extraction of their properties from experimental data to be further 
complicated by these overlapping resonances, each of which is anticipated 
to have a width of 150--300~MeV.
However, the separable nature of the model gluon propagator 
$D_{\mu\nu}(p-q)$ in Eq.~(\ref{kernel}) that leads to the exotic-vector
 meson Bethe-Salpeter  amplitude in Eq.~(\ref{Gi-+}) only supports 
{\em two} bound states in the $J^{PC}=1^{-+}$  channel.
As a result, the higher of these two bound states may be artificially
lower in the spectrum than one would expect from results from similar
calculations based on a non-separable kernel. 
The existence of higher mass exotic mesons is an issue that clearly needs
to be pursued further within more sophisticated models.

One important ramification of including odd-time parity quark-bilinear
operators (such as those considered hererin) in a study of exotic mesons
is the impact such 
operators have on the decay modes of these mesons.  In particular, it has
been argued that the decay of an exotic-$1^{-+}$ meson into a $\pi$ and
$\eta$ final state is suppressed by the Okubo-Zweig-Iuzaki rule 
\cite{Iddir}.
Such a suppression depends critically on the assumption that the structure
of the exotic meson is given by a $\bar{q}qg$ operator.  
On the other hand, if the exotic meson contained a significant mixing of
the quark-bilinear operator of the form given in Eq.~(\ref{Gi-+}), 
then such a suppression would not be expected; the exotic meson decay 
could have a significant branching ratio into the  $\pi$ $\eta$ channel.
It is interesting to note that nearly all of the decay events observed
by experiment so far are, in fact, $\pi_{1}(1400) \rightarrow \eta \pi$.
Additional studies are necessary to explore the importance of various
operators on exotic-meson decays.

Before concluding, a comment concerning the normalization of the exotic BS
amplitudes is in order.  
Historically, solutions to the homogeneous BS equation with 
exotic quantum numbers have been thought to be associated with ``negative
norm states'';
that is, the sign of the residue of the odd-time parity pole is opposite
to that of a usual meson.  
This feature of the BS equation has been explored within many model
calculations~\cite{N65}, and we observe similar behavior in the 
present study.  

The normalization condition for BS amplitudes can be derived by considering
the residue of the bound state pole appearing in the 4-point quark-antiquark
scattering amplitude.  When the BS kernel is independent of the total meson
momentum $P$, as is the case here, the normalization condition for vector (or
axialvector) mesons is given by
\begin{eqnarray}
 \lefteqn{
2 T_{\alpha\beta}(P) P_\mu \!=\! N_C \!\int\!\! 
	\frac{d^4k}{(2\pi^4)}\, {\rm tr} \bigg[ 
 } \nonumber \\  & & 
	{\Gamma}_\alpha(k,-P) \frac{\partial\;}{\partial P_{\mu}} 
	S(k + \mbox{$\frac{1}{2}$} P) 
         \Gamma_\beta(k,P) S(k - \mbox{$\frac{1}{2}$} P) 
\nonumber \\ & & 
	+{\Gamma}_\alpha(k,-P)S(k + \mbox{$\frac{1}{2}$} P) 
         \Gamma_\beta(k,P) \frac{\partial\;}{\partial P_{\mu}} 
           S(k - \mbox{$\frac{1}{2}$} P) \bigg] ,
                                    \label{norm}
\end{eqnarray}
where $N_C = 3$, $T_{\alpha\beta}(P)=\delta_{\alpha\beta} - P_\alpha
P_\beta/P^2$ is the spin-one projection operator and ``tr'' denotes a trace
over Dirac indices.  There is an analogous expression in Minkowski space. 

The result of using this canonical normalization relation, Eq.~(\ref{norm}),
is that the exotic-vector meson BS amplitude does indeed have a negative
normalization.  (The remaining states given in Table~\ref{Tab} have a
positive normalization.)  One possible ramification of a negative BS
normalization would be seen in quark-antiquark scattering amplitude, at least
in principle.  A negative BS normalization is associated with the appearence
of the pole in the quark-antiquark scattering amplitude ${\cal M}$ with a
negative residue.  However, this possibility does not immediately cause any
problems for two reasons.  
First, {\em on-shell} quark-antiquark scattering amplitudes (even if such
objects did exist) do not couple to exotic-vector mesons because the overall
factor of $q\cdot P = 0$ in Eq.~(\ref{Gi-+}) precludes such couplings to
on-shell fermions.  
Second, consequences of quantum mechanical relations such as unitarity and
the optical theorem, which could have non-trivial complications with the
appearence of negative residue states since they relate {\em off-shell}
amplitudes to each other, are here, too, avoided.  They cause no problems
since unitarity relations receive {\em no} contribution from single-particle
intermediate states \cite{PLS}.  This simple observation is due to the fact
that if the one-particle state is stable it must be lighter than any
two-particle threshold, and so is not directly produced in two-particle
scattering.  Conversely, if the one-particle state is unstable then it is
never observed asyptotically and it does not play a role in unitarity
constraints; no particle flux is carried away by unstable states.  Problems
that might arise due to exotic mesons having negative BS normalizations would
first appear in the $J^{PC}=1^{-+}$ channel of two-hadron scattering, such as
$\pi b_1$ scattering.

The real issue that must be addressed to fully determine the nature of
exotic-vector mesons obtained from the BSE is whether or not, in the absence
of open decay channels, the resulting exotic-meson state has a positive norm
in the Hilbert space.  In the present quantum field theoretic framework, this
question requires an analytic continuation of our amplitudes into Minkowski
space and then a reconstruction 
of 
the Hilbert space by means of the techniques 
and theorems developed in Wightman axiomatic field theory.  Such a program is
beyond the scope of the present study, although there is progress towards
these goals.  In particular, a procedure to carry out the first step of
explicit analytic continuation from Euclidean to Minkowski space for
amplitudes in confined-quark models is currently being successfully employed
by one of the authors.

For the moment, a complete understanding of the role of exotic mesons in the
BSE and the ramifications of a negative BS normalization remains elusive.
Still, it is our belief that any possible difficulties will be overcome in
the future since if exotic mesons do exist in nature, then there seems to be
no reason a-priori why one could not construct non-local $\bar{q} q$
interpolating fields of the form of the BS amplitude in Eq.~(\ref{Gi-+}).
Such interpolating fields could then be used to generate driving terms in
quark-antiquark scattering equations resulting in the appearence of poles in
exotic-meson channels, the observed exotic-meson states.  The existence of
such poles, in turn, imply the homogeneous BSE, Eq.~(\ref{bsv}), considered
in this study.

\section{Conclusion}
In conclusion, we have obtained masses for two $1^{-+}$-exotic mesons
from the quark-antiquark Bethe-Salpeter equation by assuming an
odd-time-parity structure of the BS amplitude.  
The masses obtained for the exotic-$1^{-+}$ mesons are 1439 and 1498~MeV.  
The mass of the lighter of the two exotics is significantly lighter than
results obtained by most other models, but seems to be in good agreement
with current experimental values. 

It is shown that the odd-time-parity amplitudes associated with these
exotic-vector mesons do indeed correspond to bound states
whose Bethe-Salpeter amplitudes have negative normalizations as found 
in earlier studies.
Nonetheless, one can argue that this by itself is not enough to disregard 
solutions of the Bethe-Salpeter equation in these channels, and
additional studies employing more realistic models are necessary before
determining the meaning of negative BS normalizations as well as any
possible pathologies that may arise from such solutions.

It is interesting that the odd-time-parity operators appearing in the
structure of exotic mesons may provide additional decay channels for
exotic mesons, such as $\pi_{1}(1400) \rightarrow \eta \pi$, that would
normally be suppressed according to the OZI rule.
This would provide a novel solution to the fact that while some models
predict an insignificant branching width for the decay 
$\pi_{1}(1400)\rightarrow \eta \pi$, experiments find this to be a
dominant decay mode.

\begin{acknowledgements}
This work is supported by the National Science Foundation under contract
PHY0071361. 
\end{acknowledgements}
%

\end{document}